# Surface Treatment of 45S5 Bio-glass using Femtosecond Laser to Achieve Superior Growth of Hydroxyapatite

S. Shaikh[1], S. Kedia[1,*], A. K. Singh[1], K. Sharma[2], S. Sinha[1]

[1]*Laser and Plasma Technology Division, Bhabha Atomic Research Centre, Mumbai 400 085, India*
[2]*Glass and Advanced Materials Division, Bhabha Atomic Research Centre, Mumbai 400 085, India*
E-mail: skedia@barc.gov.in

**Abstract:**

45S5 Hench bio-glass (BG) has gained interest in research because of its potential clinical applications. Several studies in-vivo and in-vitro have been in progress to improve bio-integration efficiency of this glass. In present contribution, surface modification of Hench BG has been done employing a femtosecond (fs) laser beam, resulting in increased effective surface area of the sample. These surface modified samples were subsequently immersed in simulated body fluid for varying number of days and characterized using Scanning electron microscope, energy dispersive X-ray analysis, X-ray diffraction, and micro-Raman spectroscopy. In-vitro studies indicated superior growth of hydroxyapatite (HAP) layer on the laser treated samples in comparison to the untreated samples. Presence of strong XRD peaks confirmed faster growth of HAP on laser treated samples. Raman peaks, five times more intense and relatively narrower represented higher crystallinity of hydroxyapatite layer on laser treated BG.

## 1. Introduction:

Use of implants made of biomaterials has emerged as a potentially successful and preferred option for treatment and repair of bones as such materials facilitate growth of calcium rich Hydroxyapatite ($Ca_{10}(PO_4)_6(OH)_2$ or HAP) layer on its surface in human plasma. Presence of HAP layer on bioactive material, in its turn, accelerates interfacial reactions between host tissues and implant materials. Therefore, extensive research has been focused on techniques which can suitably modify surface properties of implants thereby enhancing growth of HAP and hence the biocompatibility of such implants. Large varieties of bioactive materials exist; glasses, ceramics, metals, and polymers which have been used in bio clinical applications. Among these 45S5 Hench bio-glass (BG) is a well known and extensively studied bioactive material [1-2]. This BG is a subset of inorganic bioactive material and has been widely investigated as an implant material. BG has been used in more than a million patients to repair bone defects in the jaw and for orthopaedic applications, as dental inserts for tooth roots and implants restoring hearing of deaf patients [3-5]. 45S5 BG is a modified soda lime silica glass having a composition of $SiO_2$ (45 wt %), $Na_2O$ (24.5 wt %), CaO (24.5 wt %) and $P_2O_5$ (6 wt %) and can be synthesized either by, melt-casting process [6] or sol-gel method [7]. BG has been found to efficiently react with physiological fluids to form bone like HAP layer successfully enhancing biological interaction of collagen with the material surface [8].

Several in-vitro studies have been done towards improvement of growth of HAP on BG. For this, 45S5 BG has been modified either physically or chemically and immersed in simulated body fluid (SBF) to observe time dependent growth of HAP [9-11]. SBF is a mixture of NaCl, $NaHCO_3$, KCl, $Na_2HPO_4.2H_2O$, $MgCl_2.6H_2O$, $Na_2SO_4$, $(CH_2OH)_3 CNH_3$, $CaCl_2.2H_2O$ and HCl, typically a chemical equivalent to human blood plasma [12]. Growth of HAP layer on the surface of BG in SBF follows three main steps: ion exchange, dissolution and precipitation. In the initial step, cations such as, $Na^+$ and $Ca^+$ from the glass exchange with $H^+$ present in SBF. Subsequently, Si-O-Si bonds of BG break by action of hydroxyl ions ($OH^-$) and hydrated silica (SiOH) is formed. As a result, a silica-rich gel layer is generated on the surface of BG. Finally, precipitation of the calcium and phosphate ions released from the glass together with those from solution form a calcium-phosphate rich HAP layer on the surface of the glass [13].

Some groups have worked towards surface modification of BG to improve bio-integration of the sample maintaining bulk properties such as hardness and strength unchanged [14-20]. Cannillo *et al* used salt-leaching technique to produce 45S5 composite scaffolds in which well developed open interconnected porous structure was obtained [14]. Fu *et al* used polymer foam replication technique to prepare porous scaffolds of bioactive glass with high strength [15]. Song *et al* used room-temperature freeze-casting method to fabricate porous bioactive glass ceramics, leading to samples which were highly biocompatible [16]. Lopes *et al* investigated the changes that occurred on the surface of 45S5 bioactive glass when enriched with calcium ions [17]. Three-dimensional highly porous bioactive scaffolds were fabricated by sintering process by Chen *et al* after which the samples showed good mechanical support and bioactivity [18]. Some of the surface modification techniques discussed above required use of organic solvents which however, reduces ability of cells to form new tissues in vivo [19]. High temperature when used for sintering process crystallizes the BG sample which consequently reduces its bioactivity [20]. Therefore, there exists a need for a technique which can modify the surface of BG samples and eventually improve its bioactivity.

We report here, use of a novel laser based approach employing laser induced surface modification of BG resulting in increased roughness and hence an increase in the effective surface area of the sample. We have used a direct femtosecond laser writing technique to surface modify 45S5 BG samples. Laser based process being a one-step method, not requiring any organic solvent, also allows processing to be performed under ambient conditions. In addition, unique advantages associated with femtosecond laser based surface treatment, such as reduced debris and contamination, excellent reproducibility and minimum heat affected zone, make this approach a particularly attractive technique [21]. Appropriate focussing of the processing femtosecond laser beam allows modification either, on the surface or, within the bulk of the material. In our investigation, precisely focused laser beam on the sample surface restricted the modification only to the surface of the sample. Parameters such as laser power and sample scanning speed were also suitably optimized after several runs leading to surface modification of samples, at the same time avoiding sample damage by the intense laser beam. Subsequently, these laser treated BG samples were immersed in SBF for varying number of days to investigate development of HAP. Our results demonstrated far superior growth of HAP on the laser treated BG samples in comparison to untreated samples.

## 2. Experiment:

### *2.1. Preparation of bio-glass:*

45S5 glass with nominal composition $SiO_2$ (45 wt %), $Na_2O$ (24.5 wt %), CaO (24.5 wt %), and $P_2O_5$ (6 wt %) was prepared by melt-quench process. Around 100g glass batch was prepared by mixing reagent grade $SiO_2$, $CaCO_3$, $NH_4H_2PO_4$ and $Na_2CO_3$. The charge was calcined at a maximum temperature of 900$^o$C for 12 h, holding at intermediate temperature for 6-8 h, decided by the decomposition temperatures of various precursors. To ensure complete decomposition of carbonates (precursors) into oxides, the batch was weighed before and after calcination of the precursors. The calcined charge was melted under air ambient at around 1500°C in a Pt-Rh crucible. The melt was poured in graphite mould and annealed at around 500°C for 4h. The fabricated glass was cylindrical in shape which was cut into 2 mm thick circular disks having a diameter of 1.5 cm. Flat surfaces of the sample were polished and laser treatment was performed on the polished surface.

### *2.2. Femtosecond laser based surface modification:*

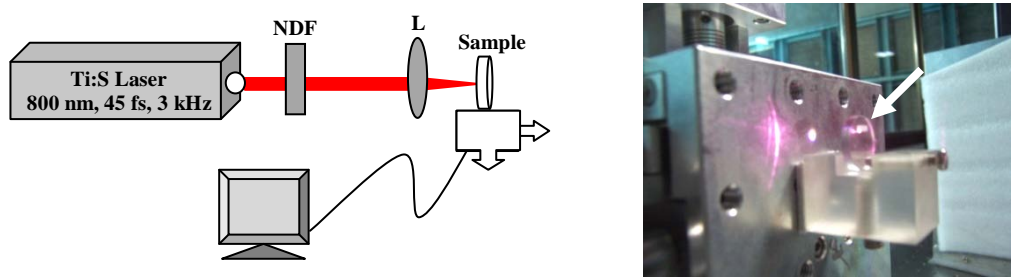

*Fig. 1: (a) Illustration of experimental arrangement used for laser treatment of BG NDF is neutral density filter, L is 5 cm focal length lens, and sample was fixed on a computer controlled XY-stage, (b) Photograph of the BG during surface texturing process, the laser treated area on the BG is indicated with an arrow*

For surface modification, beam from a pulsed Titanium : Sapphire (Ti:S) laser at a wavelength of 800 nm with pulse width 45 *fs* and repetition rate 3kHz was focused on to the surface of the glass sample using a 5 cm focal length lens, as shown in Fig. 1a. The surface modification or laser writing speed was controlled using a computer controlled XY-translational stage on which the BG sample was mounted. Average laser power was varied between 50 to 200 mW and various sample scanning speeds were tried to arrive at a condition where microstructures could be generated without formation of crater and associated damages on the surface of BG samples. Typical values of optimized average laser power and scanning speed employed for surface treatment were 200 mW and 20 µm/sec, respectively. This corresponds to an average laser fluence of 1 J/cm$^2$ incident on the BG sample. A photograph of the sample during laser writing process is shown in Fig. 1b. The laser treated area on the sample is indicated using an arrow. These laser treated samples were subsequently immersed in SBF for varying number of days such as, 1 day, 3 days, 5 days, 10 days and 20 days to investigate the extent and rate of HAP growth.

### 2.3. In vitro bioactivity tests:

Simulated body fluid was used to evaluate in vitro bioactivity of the samples. SBF has an ionic composition similar to that of human blood plasma. The SBF solution was prepared by dissolving appropriate amount of salts (as listed in Table-I), in de-ionized water. 1000ml of SBF was prepared in which the salts were initially dissolved in 700ml of water and a total of 40ml of 1M HCl solution was consumed for pH adjustments. The reagents were added one by one (in the order provided in Table-I) after each was completely dissolved in water. 15ml of HCl was added just before 6$^{th}$ reagent addition. The remaining portion of HCl solution was utilized in further titration process. The prepared solution was finally diluted with de-ionized water to make the final volume of 1L.

Femtosecond laser treated and untreated glass samples were immersed in SBF at body temperature (37°C) and the pH of SBF was maintained at 7.4, similar to that of blood. Freshly prepared SBF solution was used each day to ensure an availability of fresh reactive ions for the reaction to occur with samples. At the end of the soaking step for BG, the samples were removed from the SBF and rinsed with deionised water to prevent further reactions and dried in air before carrying out the characterization tests.

| Order | Reagent | Amount (gpl) |
|---|---|---|
| 1 | NaCl | 6.547 |
| 2 | $NaHCO_3$ | 2.268 |
| 3 | KCl | 0.373 |
| 4 | $Na_2HPO_4 \cdot 2H_2O$ | 0.178 |
| 5 | $MgCl_2 \cdot 6H_2O$ | 0.305 |
| 6 | $CaCl_2 \cdot 2H_2O$ | 0.368 |
| 7 | $Na_2SO_4$ | 0.071 |
| 8 | $(CH_2OH)_3CNH_3$ | 6.057 |

Table I. Chemical composition of SBF

*2.4. Characterization:*

The surface morphology of BG before and after dipping in SBF was investigated using a scanning electron microscope (SEM) (Carl Zeiss EVO 40 SEM, 20 KeV beam energy, Tungsten filament). An elemental distribution analysis was carried out to identify the ionic substitution route for growth of HAP using an energy dispersive X-ray analyzer (EDAX) (Bruker Quanta EDS), which was directly connected to a SEM system. Surface roughness of the samples was measured using a Taylor Hobson 3D optical profilometer (Green light, Resolution 0.1Å). Crystalline nature of HAP grown on glass surface was confirmed from X-ray diffraction (XRD) studies using PANalytical MRD system. XRD measurements were done using CuKα radiation of wavelength 1.54 A$^o$ in out of plane geometry over the 2θ range of $20^0$-$35^0$ at scanning speed of $2^o$/min. Micro-Raman spectroscopic investigation was performed using HR 460 spectrograph, and excitation source at 532 nm from a diode- pumped solid-state laser. Liquid nitrogen-cooled CCD detector and an appropriate super-notch filter to cut off the Rayleigh scattered light were used. Signals were collected in the back-scattering geometry and Neon lines were used for calibration of Raman spectra.

## 3. Results and discussion:

### *3.1 Effect of laser treatment on 45S5 BG:*

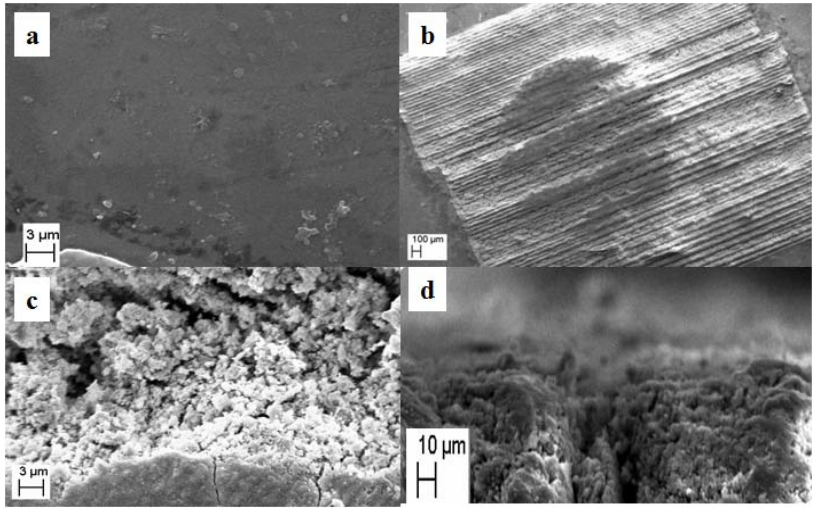

*Fig 2: SEM images of (a) surface of untreated BG, (b) & (c) Typical low (46X) and high (3KX) magnification SEM images of laser treated region of BG and (d) cross-sectional view of laser treated BG*

Fig. 2a is a typical SEM image of the untreated BG surface. In Figs 2b and 2c are shown SEM images of the laser treated BG at low and high magnification levels. An area of 4mmX5mm was surface modified by femtosecond laser as depicted in Fig. 2b. Microstructures formed on the surface of the glass along with increased surface roughness after laser treatment is clearly visible in Fig. 2c. In comparison a relatively smoother surface was observed in case of an untreated sample. Extent of laser modification of BG surface was found to be hundreds of micron in depth which was far less than the total thickness of the sample as shown in the cross-sectional view of the laser treated sample (Fig 2d).

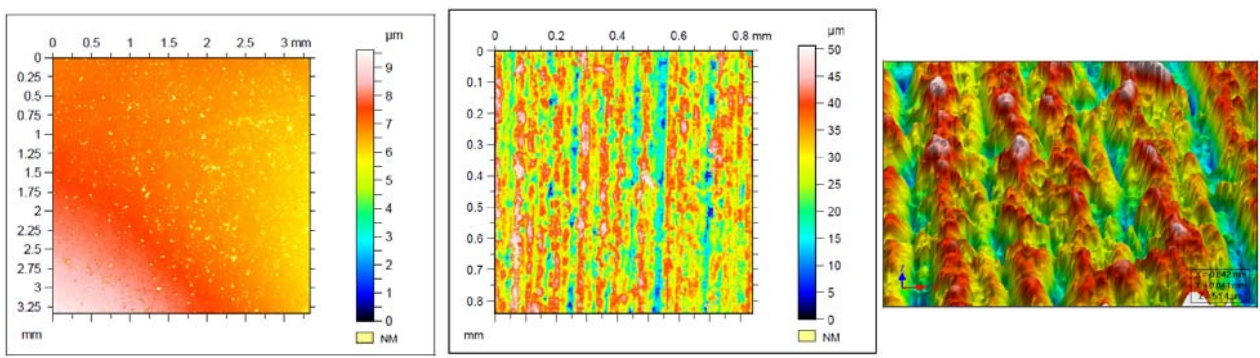

*Fig 3: Optical images of BG (a) untreated, laser treated (b) 2D, and (c) 3D*

Surface roughness of untreated and laser treated BG was measured using optical profilometer. Fig. 3a is the optical image of untreated BG, the average surface roughness of the sample was measured as 0.23 μm. Figs. 3b and 3c are the 2D and 3D optical images of laser treated sample, respectively. The surface roughness of the BG notably increased to 6.41 μm after laser treatment.

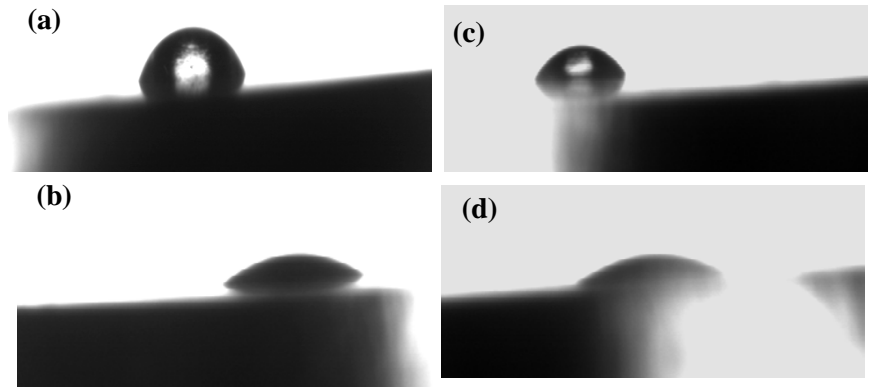

*Fig 4: Water contact angle of (a) untreated (b) laser treated, SBF contact angle of (c) untreated and (d) laser treated, BG*

While, as is samples were hydrophilic for both water and SBF, their wettability is expected to be modified post laser treatment. Water contact angle of BG changed from $80^o$ (Fig. 4a) to $34^o$ (Fig. 4b) for untreated and laser treated samples, respectively. The SBF drop contact angle also decreased from $58^o$ (Fig. 4c) to $30^o$ (Fig. 4d) after laser treatment. The improved wettability of the laser treated surface for SBF is expected to facilitate increased polar interaction between sample and SBF hence boosting the bio-integration process.

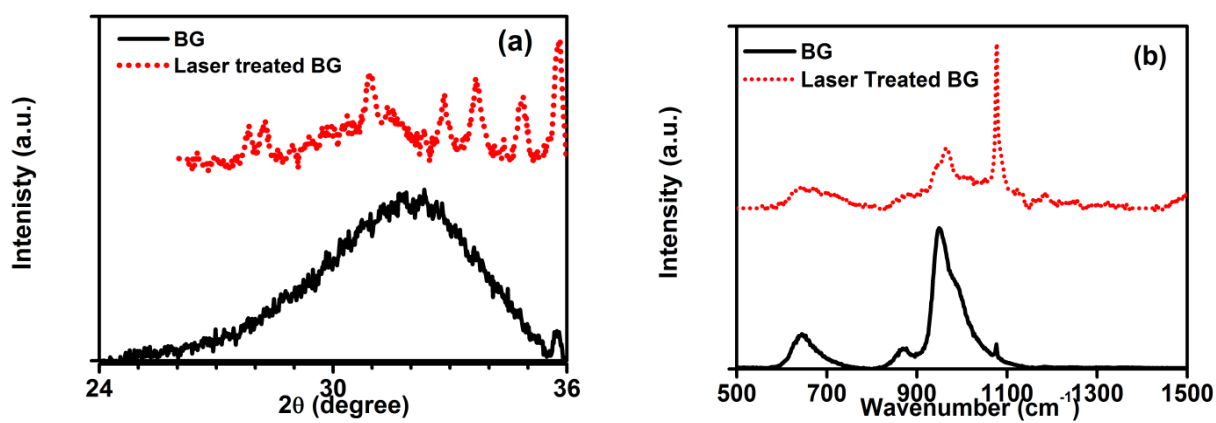

*Fig 5: (a) X-ray diffraction pattern of untreated (solid line) and laser treated (dotted line) BG, and (b) Raman spectra of untreated (solid line) and laser treated (dotted line) BG*

Solid and dotted lines in Fig. 5a are the XRD patterns of untreated and laser treated BG, respectively. The broad solid line centred at 31.8° belongs to as produced BG confirming the amorphous nature of the glass. A number of sharp peaks are observed in XRD pattern on laser treatment, suggesting surface crystallization of the sample on laser irradiation. Most of the XRD peaks in Fig. 5a matched with JCPDS # 22.1455 confirming $Na_2Ca_2Si_3O_9$ crystalline phase of the BG. Similar phase change in the BG was reported by Chen *et al* when the glass was sintered at a high temperature of 1000° C [18]. Our observations therefore suggested that, during laser treatment heat accumulated within the focal volume of the laser treated zone was sufficient to generate some of these crystalline phases. Presence of crystalline phase of the BG has been reported to delay the onset time of HAP growth on BG hence decreasing its surface reactivity [22]. However, phase change observed by us on laser treatment was restricted within the laser affected zones (thin layer) on the sample surface leaving the bulk largely unaffected. As depicted in Fig.2d, laser affected thickness is restricted to a few hundreds of micron in comparison to total thickness of the glass sample (2000 micron). In contrast to surface crystallization, increased surface roughness generated on laser treatment is expected to enhance bioactivity. Since, the crystallization of the sample has occurred in micron-level, this is not expected to adversely affect the bio-activity of glass significantly.

In Fig. 5b solid and dotted lines show the Raman spectra of as prepared and laser treated BG, respectively. The Raman bands observed for BG mainly originate from the vibrations of bond in $SiO_4$ tetrahedra and Si-O-Si bonds linking the tetrahedral. Peaks located at wavenumbers greater than 800 $cm^{-1}$ have been attributed to symmetric stretching vibration of $SiO_4$ tetrahedra with different numbers of non-bridging oxygen (NBO) and peaks below 800 $cm^{-1}$ indicates bending vibrations of Si-NBO bonds. The broad band centred at 645 $cm^{-1}$ corresponds to Si-O-Si groups, peak at 870 $cm^{-1}$ is associated with monomers $SiO_4^{4-}$ (4NBO), the most intense peak centred at 946 $cm^{-1}$ is assigned to dimers $Si_2O_7^{6-}$ (3NBO) and peak at 970 $cm^{-1}$ ascribes the rings and chains $Si_2O_6^{4-}$ (2 NBO) [23-24]. Finally, the peak at 1076 $cm^{-1}$ is again assigned to vibrational modes of Si-O-Si [25]. Subsequent to laser treatment, the peak associated with dimer group shifted from 946 $cm^{-1}$ to 965 $cm^{-1}$ and its peak intensity reduced significantly, indicating reduction in the $Si_2O_7^{6-}$ (3NBO) bond strength in the sample. Noticeable enhancement in the peak intensity of the Si-O-Si band located at 1076 $cm^{-1}$ confirmed strong crystallization of silicate glass after laser treatment.

**3.2 In-vitro studies after laser treatment:**

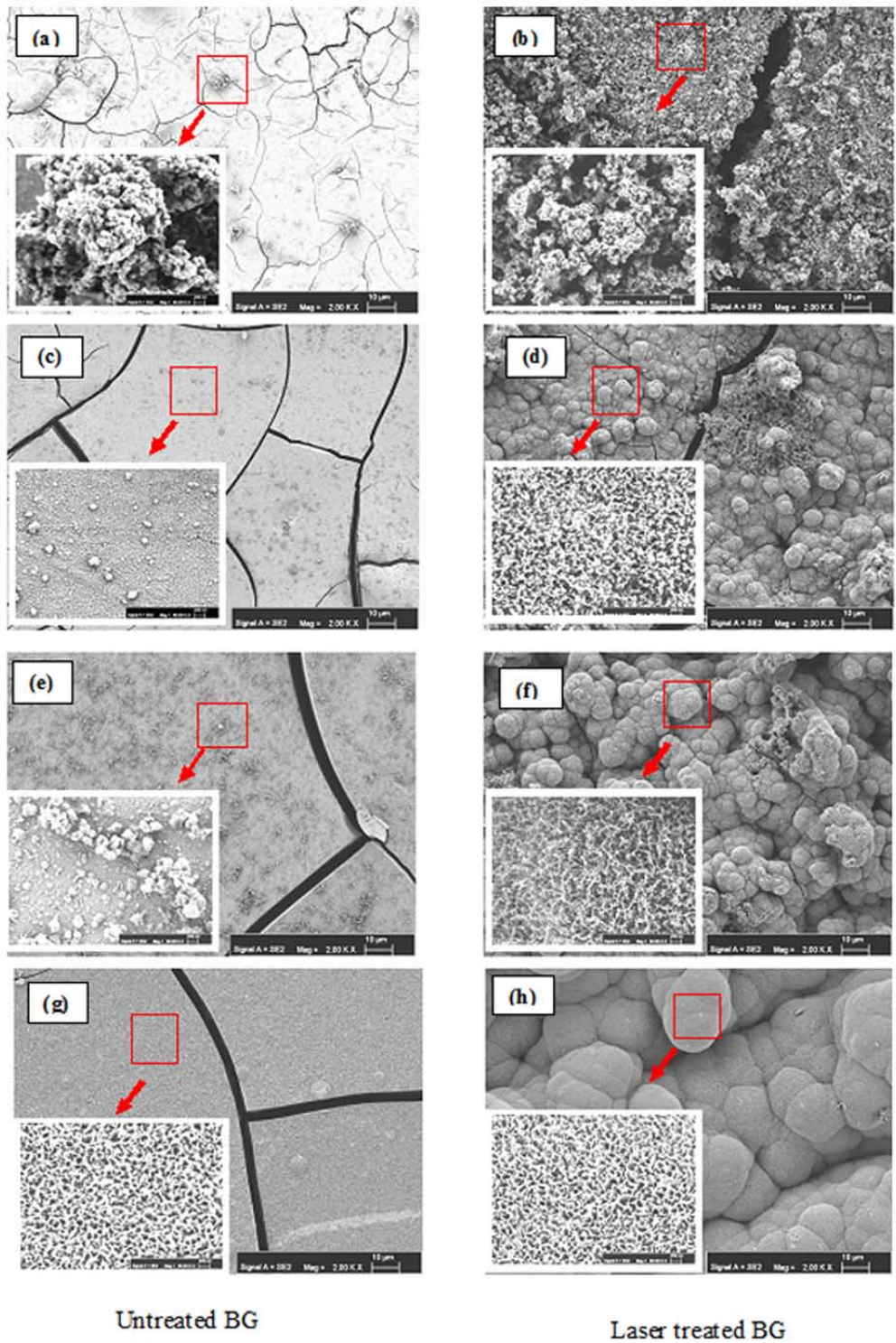

*Fig 6: SEM images of untreated and laser treated BG immersed in SBF for (a-b) 3 days, (c-d) 5 days, (e-f) 10 days and (g-h) 20 days [scale bar 10 μm], inset shows magnified SEM images [scale bar 200 nm] of corresponding sample*

In Fig. 6 are shown SEM images of untreated (left) and laser treated (right) BG that had been dipped in SBF for varying number of days. After 3 days in SBF, nucleation sites of HAP developed in regions separated far apart from each other on the un-treated BG surface (Fig. 6a). For the same period of time, relatively denser nucleations of HAP were seen on the laser treated BG (Fig. 6b). Nucleation density increased on untreated glass samples after being dipped in SBF for 5 days and 10 days (Fig. 6c & 6e). However, in case of laser treated BG spherical agglomeration of HAP completely covering the entire laser treated region of BG was observed with increasing number of days (Fig. 6d & 6f). Similar morphology of HAP has been reported by Song *et al* [16] and Chen *et al* [18] on the surface of BG that had been sintered at $1000^0$ C. The spherical structure of HAP on the laser treated area shows a network of thread like structure when magnified, shown in the inset of Fig. 6d and 6f. Such structures were not observed on untreated BG prior to 20 days of immersion. After 20 days in SBF, the untreated BG shows a more uniform growth of HAP (Fig. 6g) and for the same number of days high quality and dense HAP growth was observed on laser treated sample, as is evident in Fig. 6h. The agglomeration rate and density of thread like structure were found to markedly increase when samples were laser treated and dipped in SBF. That the growth of HAP is thicker on the laser treated BG surface is also clearly evident from the fact that cracks generated during synthesis of BG continue to remain clearly visible in the SEM images of untreated BG after HAP growth (Fig. 6g) unlike in case of laser treated samples where these cracks get effectively filled up by HAP growth, as seen in Fig. 6h. Further, the laser treated samples developed HAP layer with spherical features of increasing size when immersed in SBF, whereas, such features were hardly observed on untreated samples. Typical diameter of these spherical agglomerates for 5days, 10days and 20days of immersion period in SBF were found to be approximately 5µm, 10µm and 20µm respectively.

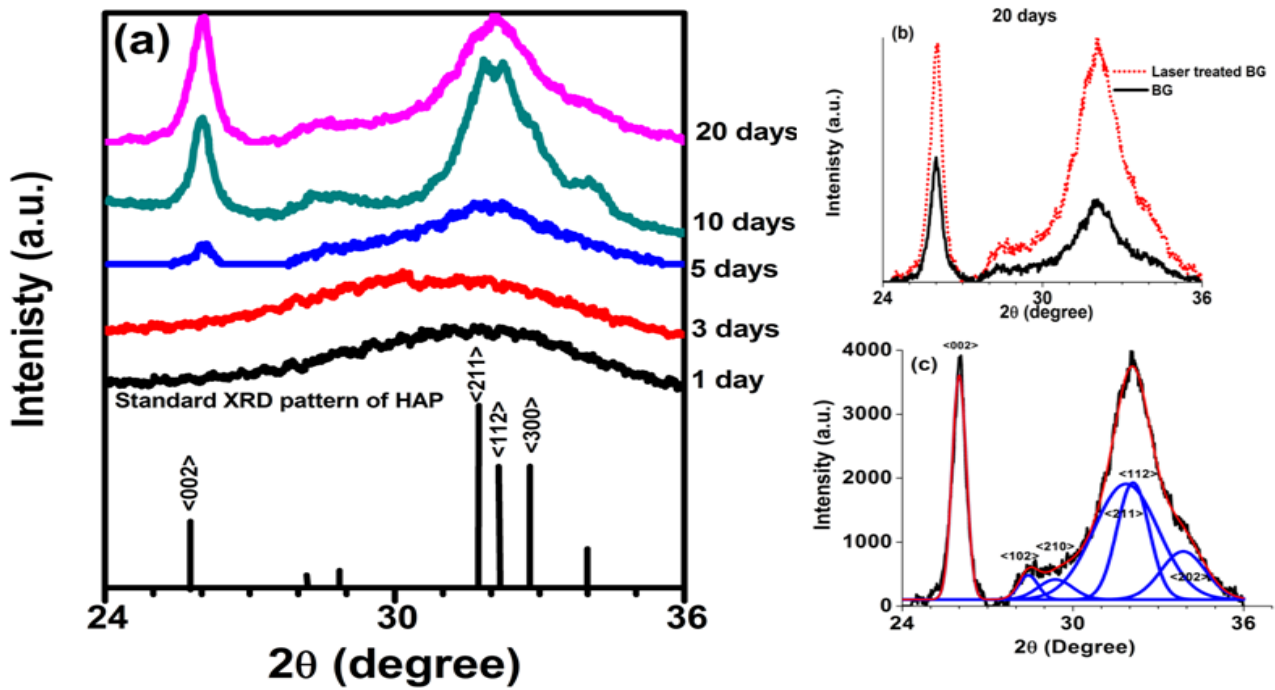

*Fig 7 (a) X-ray diffraction pattern of laser treated BG dipped in SBF for different number of days, & Standard XRD pattern of HAP (bar graph) (b) XRD of untreated (solid line) and laser treated (dotted line) BG immersed in SBF for 20 days, and (c) deconvolution of XRD plot of laser treated BG immersed in SBF for 20 days.*

Fig 7a shows the XRD patterns of laser treated BG immersed in SBF for varying number of days along with the standard XRD pattern of HAP (JCPDS # 09-0432). HAP peaks were not detected for the samples dipped in SBF for 1 and 3 days. That growth of HAP was not uniform till 3 days as observed in Fig. 7b could probably explain this absence of XRD peaks corresponding to HAP in Fig. 7a. However, multiple XRD peaks which were observed arising due to crystallization of glass after laser treatment (dotted curve in Fig. 5a) also got submerged in the broad band pattern seen in Fig. 7a for these two cases (1 day and 3 days). This could happen on account of reaction of micro-crystalline BG with SBF [4]. XRD peaks which were signature of micro crystalline BG disappeared faster (within 1 day) in our study in comparison to work reported in Ref-4 (3-days). After 5 days, a continuous layer of HAP formed and covered the laser treated surface and therefore, a peak at $26^o$ along with a broad trace centred at angular positions $32^0$ are observed in XRD. As confirmed from JCPDS # 09-0432, these peaks belong to <002> and <112> planes of HAP. For laser treated BG samples which had been dipped for 10 and 20 days, these peaks became sharper with decreased spectral width, and increased intensity signifying improvement in the crystalline phase of HAP. Typically, the standard ratio of peaks corresponding to the <002> and <112> planes of HAP is 0.66 (JCPDS #09-0432). However our observations indicate peaks representing these planes to be of nearly equal intensity. This suggested that in our case, the HAP crystals are randomly oriented with respect to glass surface [26]. Fig 7b shows XRD pattern of untreated (solid line) and laser treated (dotted line) BG samples both of which had been dipped in SBF for 20 days. The peak intensities corresponding to <002> and <112> planes in case of laser treated BG far exceeds those corresponding to untreated glass, implying faster growth of HAP after laser treatment. As observed in Fig. 7a, the micro-crystalline top layer of BG dissolved in SBF within 1-day leaving behind surface microstructures which served as nucleation sites aiding initial reaction of BG surface with SBF. This was also confirmed by comparing Fig 6a with Fig. 6b where denser HAP growth was seen on laser treated BG than on surface of untreated BG. Therefore our observations confirmed that faster growth of HAP occurred on laser treated sample in comparison to untreated BG for the same immersion time. Further, XRD pattern of laser treated BG immersed in SBF for 20 days was deconvoluted (Fig. 7c) and the resultant peaks matched well with the data available in standard JCPDS (09-0432) for crystalline HAP. The average HAP grain size as determined using Debye-Scherrer formula typically ranged between 17nm-25nm.

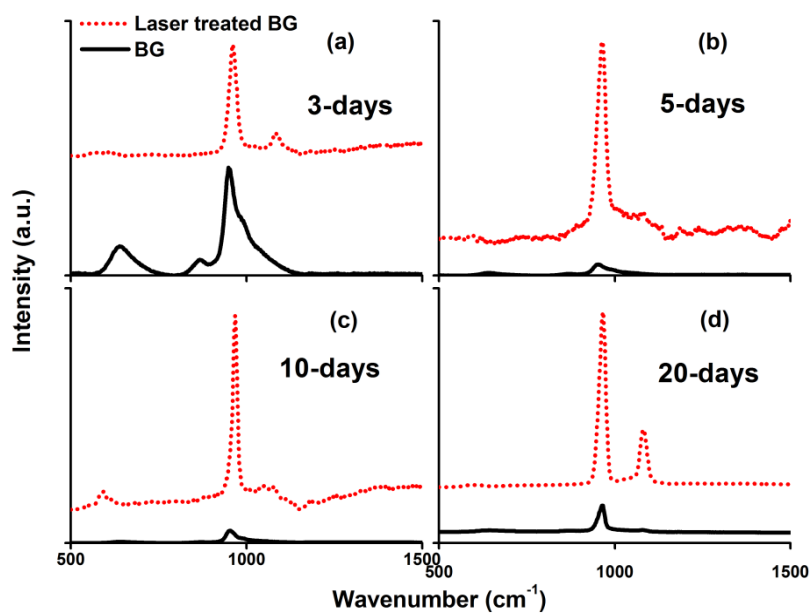

*Fig 8: Micro-Raman spectra of untreated (solid line) and laser treated (dotted line) BG dipped in SBF for (a) 3 days, (b) 5 days, (c) 10 days and (d) 20 days*

To determine the various vibrational modes and associated phases present in HAP grown on BG, Raman spectroscopy was performed on the samples dipped in SBF for varying number of days and results are shown using solid (untreated BG) and dotted (laser treated BG) lines in Fig. 8. Analysis of observed Raman peaks suggested that these correspond to vibrational modes of $PO_4$ group associated with HAP. The strongest peak at 965 cm$^{-1}$ is assigned to the symmetric stretching mode of $PO_4$ tetrahedron. This phosphate symmetric stretch peak is the characteristic peak of HAP, shifting of this peak towards higher wavenumber indicates [27] higher crystallization of HAP particles. The second intense peak at 1078 cm$^{-1}$ can be assigned to the stretching-vibration $CO_3^{2-}$ (carbonate) bond [27]. Percentage of carbonate ion in HAP mainly depends upon the time for which the sample has been dipped in SBF, the intensity of this peak increases as carbonate content in HAP increases [28-29]. Significant increase in the intensity along with noticeable reduction in spectral width of the phosphate symmetric stretch peak centred at 965 cm$^{-1}$ indicates sizeable growth and greater crystallinity [30] of HAP on laser treated BG surface. Up to 5 times enhancement in the Raman intensity was observed for laser treated BG in comparison to untreated BG when dipped in SBF for 20 days (Fig. 8d). The Raman intensity of carbonate peak located at 1078 cm$^{-1}$ was also found to be greater in case of laser treated sample.

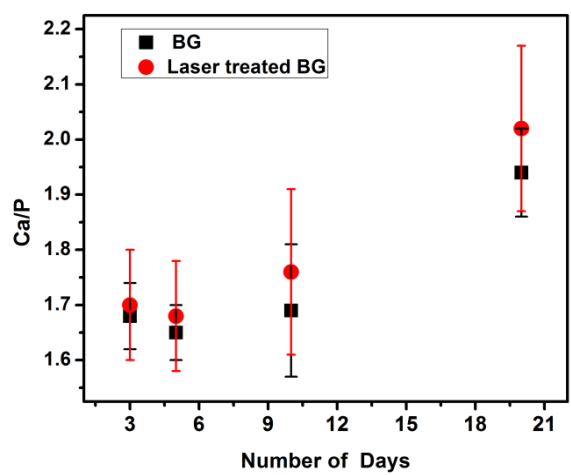

*Fig 9: Ca/P ratio of HAP grown on the surface of untreated (squares) and laser treated (circles) BG with respect to number of days*

Elemental composition and nature of HAP grown on the BG was also analyzed on the basis of EDX spectroscopy and calcium to phosphorus (Ca/P) ratio as determined through EDX analysis. The Ca/P ratio of HAP on untreated (squares) and laser treated (circles) BG immersed in SBF for varying number of days are shown in Fig. 9. The Ca/P ratio of HAP on laser treated samples was found to be marginally higher, though largely comparable to untreated samples. Observed value of Ca/P atomic ratio was higher than the ideal stoichiometric value of 1.67 for pure HAP. Higher calcium to phosphorus atomic ratio has been reported for samples with carbonate ($CO_3^{2-}$) group substitution into the apatite [31]. This is consistent with our micro-Raman spectroscopic analysis (Fig. 8) which also indicated enhanced presence of carbonate ion in HAP on the laser treated BG samples when dipped in SBF. Presence of carbonate ions in HAP has been of potential interest as it has been associated with higher mechanical strength and has also been reported to play a crucial role in bone metabolism through higher solubility of apatite [31].

The main concerns of HAP development on implants are its physical quality broadly characterized by uniformity and crack free nature of the coating and its high crystalline phase [26]. Uniform growth of HAP enhances its adhesion with the host and crystalline nature increases its strength and bond to bone tissues. Both these requirements are addressed by laser surface treatment of BG samples as confirmed in our present investigation. While, SEM images (Fig. 6d, 6f and 6h) confirmed uniform and continuous growth of HAP completely covering the laser treated area, XRD and Raman results established growth of highly crystalline HAP on laser treated BG.

In conclusion, controlled surface modification of 45S5 BG was successfully performed in ambient atmosphere employing a femtosecond laser. SEM images of the laser treated surface confirmed generation of microstructures on the surface of BG samples on laser treatment which enhances the roughness and hence the effective surface area of the sample. This facilitated rapid interaction of the BG surface with SBF leading to faster and uniform growth of HAP. Laser surface treatment thus can serve as a potential technique enabling rapid growth of good quality HAP coating on this type of BG samples. In comparison to techniques such as sintering of BG at high temperature for improved bio-integration, laser based surface treatment offers a superior alternative allowing targeted surface modification without altering bulk properties such as porosity and hence its mechanical strength. Our EDX and micro Raman spectroscopy results indicate the presence of carbonate ions in HAP grown on laser treated surface which also is advantageous as it is known to control bone metabolism. Our XRD characterization indicated growth of randomly oriented crystalline HAP domains on laser treated BG and Raman analysis confirmed presence of relatively intense characteristic peak of HAP on laser treated samples. All our results on laser treated BG samples consistently demonstrated growth of superior HAP in comparison to untreated BG. Hence, controlled surface modification using femtosecond laser pulse opens up new possibilities to enhance the bio-activity of the Hench BG while keeping the bulk properties unchanged.

**Acknowledgment:**

Authors acknowledge Dr. S. Basu from SSPD, Mr. A. Chitnis, Dr. N. Garg from HP&SRPD, Dr. A. K. Sahu from G&AMD, Dr. A. K. Chauhan from TPD, Dr. K. Singh from MSD, all from BARC, for their technical support.

**Reference:**

[1] Jones JR. Review of bioactive glass: from Hench to hybrids. Acta Biomater 2013; 9(1):4457-86.

[2] Hench LL. The story of bioglass. Journal of Material Science in Medicine 2006; 17(11): 967-978.

[3] Hench LL, Fenn MB, and Jones JR. Clinical Applications of Bioactive Glasses for Maxillofacial Repair. World Scientific. Singapore 2011; 77–96.

[4] Rust, Kevin R, Singleton, George T, Wilson, June Antonelli, Patrick J. Bioglass Middle Ear Prosthesis: Long-Term Results. American Journal of Otology 1996; 17(3):371-4.

[5] Stanley HR, Hall MB, Clark AE, King CJ, Hench LL, Berte JJ. Using 45S5 bioglass cones as end osseous ridge maintenance implants to prevent alveolar ridge resorption: a 5-year evaluation. The International Journal of Oral & Maxillofacial Implants 1997; 12(1):95-105.

[6] Pilar Sepulveda, Julian R. Jones and Larry L. Hench. Characterization of melt-derived 45S5 and sol-gel–derived 58S bioactive glasses. Journal of Biomedical Materials Research 2001; 58(6):734-740.

[7] Hamidreza P and Nychka JA. Sol–Gel Synthesis of Bioactive Glass-Ceramic 45S5 and its in vitro Dissolution and Mineralization Behavior. J. Am. Ceram. Soc 2013; 96(5):1643–1650.

[8] Hench LL. Bioceramics. J. Am. Ceram. Soc 1998; 81(7):1705–28.

[9] Hassan MM, Moawad Z, and H.Jain. Creation of Nano–Macro-Interconnected Porosity in a Bioactive Glass–Ceramic by the Melt-Quench-Heat-Etch Method. J.Am.Ceram.Soc 2007; 90 (6):1934–1936.

[10] Adams LA, Essien ER, Shaibu RO, Oki A. Sol-Gel Synthesis of $SiO_2$-CaO-$Na_2O$-$P_2O_5$ Bioactive Glass Ceramic from Sodium Metasilicate. New Journal of Glass and Ceramics 2013; 3: 11-15.

[11] Mona OA, Abdelghany AM, E.Hatem A. Corrosion mechanism and bioactivity of borate glasses analogue to Hench's bioglass. Processing and Application of Ceramics 2012; 6(3): 141-149.

[12] Tadashi Kokubo Hiroaki Takadama. How useful is SBF in predicting in vivo bone bioactivity? Biomaterials 2006; 27(15):2907-15.

[13] Lefebvre L, Chevalier J, Gremillard J, Zenati R, Thollet G, Bernache-Assolant D, Govin A. Structural transformations of bioactive glass 45S5 with thermal treatments. Acta Biomater 2007; 55: 3305–3313.

[14] Cannillo V, Chiellini F, Fabbri P, Sola A. Production of Bioglass® 45S5 – Polycaprolactone composite scaffolds via salt-leaching. Composite Structures 2010; 92(8):1823–1832.

[15] Fu Q, Rahaman MN, Bal BS, Brown RF, Day DE. Mechanical and in vitro performance of 13–93 bioactive glass scaffolds prepared by a polymer foam replication technique. Acta Biomaterialia 2008; 4 (6): 854–1864.

[16] Song JH, Koh YH, Kim HE, Li LH, Bahn HJ. Fabrication of a Porous Bioactive Glass–Ceramic Using Room-Temperature Freeze Casting. Journal of the American Ceramic Society 2006; 89(8): 2649–2653.

[17] Lopes JH, Mazali IO, Landers R, and Bertran CA. Structural Investigation of the Surface of Bioglass 45S5 Enriched with Calcium Ions. J. Am. Ceram. Soc 2013; 96 (5):1464–1469.

[18] Chen QZ, Thompson ID, Boccaccini AR. 45S5 Bioglass®-derived glass–ceramic scaffolds for bone tissue engineering. Biomaterials 2006;27( 11):2414–2425.

[19] Antonios G, Mikos, Johnna S. Temenoff. Formation of highly porous biodegradable scaffolds for tissue engineering. Biotechnology of human disorders 2000; 3(2):0717-3458

[20] Bellucci D, Cannillo V, Sola A. An overview of the effects of thermal processing on bioactive glasses. Science of Sintering 2010; 42(3): 307-320.

[21] Krol DM. Femtosecond laser modification of glass. Journal of Non-Crystalline Solids 2008; 354 (2–9):416–424.

[22] Guy PL, Hench LL. Effect of crystallization on apatite-layer formation of bioactive glass 45%.Journal of Biomedical Materials Research 1996; 30:509-514.

[23] Ziemath EC and Aegerter MA. Raman and infrared investigations of glass and glass-ceramics with composition $2Na_2O \cdot 1CaO \cdot 3SiO_2$. Journal of Materials Research 1994; 9 (01): 216-225.

[24] Bellucci D, Bolelli G, Cannillo V, Cattini A, Sola A. In situ Raman spectroscopy investigation of bioactive glass reactivity: Simulated body fluid solution vs. TRIS-buffered solution. Materials Characterization 2011; 62:(10)1021–1028.

[25] Zheng K, Solodovnyk A, Li GW,O.M., Stahli C, Nazhat S N, and Boccaccini AR. Aging Time and Temperature Effects on the Structure and Bioactivity of Gel-Derived 45S5 Glass-Ceramics. J. Am. Ceram. Soc 2015; 98(1):30-38.

[26] Liu D, Savino K, Yates MZ. Coating of hydroxyapatite films on metal substrates by seeded hydrothermal deposition. Surface 2011; 205(16):3975–3986.

[27] Hong Z, Luan L, Paik SB, Deng B, Ellis DE, Ketterson JB, Mello A, Eon JG, Terra J, and Rossi A. Crystalline hydroxyapatite thin films produced at room temperature- An opposing radio frequency magnetron sputtering approach. Thin Solid Films 2007; 515(17):6773-6780.

[28] Khan AF, AF Wais M, Khan AS, Tabassum S, Chaudhry AA & Rehman IU. Raman Spectroscopy of Natural Bone and Synthetic Appetites. Applied Spectroscopy Reviews 2013; 48(4):329-355.

[29] Balamurugan A, Michel J, Fauré J, Benhayoune H, Wortham L, Sockalingum G, Banchet V, Bouthors S, Laurent-Maquin D, Balossier G. synthesis and structural analysis of sol gel derived stoichiometric monophasic hydroxyapatite. Ceramics − Silikáty 2006;50 (1) :27-31.

[30] Freeman JJ, Wopenka B, Silva MJ, Pasteris JD. Raman Spectroscopic Detection of Changes in Bioapatite in Mouse Femora as a Function of Age and In Vitro Fluoride Treatment. Calcified Tissue International 2001; 68(3): 156-162.

[31] Murugan R, Ramakrishna S. Production of ultra-fine bioresorbable carbonated hydroxyapatite. Acta Biomaterialia 2006; 2(2):201–206.